\documentclass[fp,twocolumn]{jpsj3}
\usepackage{txfonts}
\usepackage{graphicx}

\title{Entanglement Entropy on the Boundary of the Square-Lattice $\pm J$ Ising Model}

\author{Yoshinori Sasagawa$^{1,2}$, Hiroshi Ueda$^{3,4}$, Jozef Genzor$^{1,5}$, 
Andrej Gendiar$^6$, and Tomotoshi Nishino$^1$\thanks{nishino@kobe-u.ac.jp}}
\inst{
$^1$Department of Physics, Graduate School of Science, Kobe University, Kobe 657-8501, Japan \\
$^2$Sysmex Corporation, Kobe, Hyogo 651-0073, Japan \\
$^3$Computational Materials Science Research Team, RIKEN Center for Computational Science 
(R-CCS), Kobe, Hyogo 650-0047, Japan \\
$^4$JST, PRESTO, Kawaguchi, Saitama 332-0012, Japan \\
$^5$Department of Physics, National Taiwan University, 
Taipei 10617, Taiwan \\
$^6$Institute of Physics, Slovak Academy of Sciences, 
Bratislava 845 11, Slovakia}

\abst{
The entanglement entropy $S$ is calculated on the system boundary of the 
square-lattice $\pm J$ Ising model by the time-evolving block decimation (TEBD) 
method. The random average $\langle S \rangle$ is evaluated on the Nishimori line, 
through the successive multiplications of transfer matrices, whose width $N$ is up to $300$. 
It is confirmed that $\langle S \rangle$ shows critical singularity around the Nishimori point. }

\begin{document}
\maketitle

\section{Introduction}

The effect of randomness on magnetic phenomena has been one of the issues in statistical physics. 
A well-investigated system is the Edwards--Anderson model~\cite{Edwards}, 
which was introduced for the analysis of the Mn-Cu alloy~\cite{Cannela}. 
A special case is the $\pm J$ Ising model, where neighboring interactions choose $-J < 0$ 
or $J > 0$ randomly. In two dimensions, the model exhibits either ferromagnetic or paramagnetic 
states~\cite{Morgenstern, Hasenbusch}, when there are only short-range interactions. The 
spin-glass state appears in higher dimensions or in the presence of long-range 
interactions~\cite{Bhatt}.
The analytic form of the internal energy can be obtained for the $\pm J$ Ising model, when the 
temperature $T$ and the probability $p$ of finding a ferromagnetic bond satisfy the so-called 
Nishimori condition~\cite{Nishimori1, Nishimori2}, which is represented by a curve known as the 
{\it Nishimori line} in the parameter space. 

We consider the square-lattice $\pm J$ Ising model on the finite-size lattice of width $N$, 
and perform numerical analysis using the transfer matrix 
formalism~\cite{Ozeki, Aarao, Honecker1, Merz, Honecker2}. Since the dimension of the 
matrix increases exponentially with $N$, direct numerical treatment is limited up to 
$N \sim 30$ at most. Merz and Chalker introduced a fermionic representation and extended 
the size up to $N = 256$, while most of the numerical data were collected up to 
$N = 64$~\cite{Merz}. To treat larger systems, we employ the time-evolving block 
decimation (TEBD) method~\cite{Vidal, Daley, Verstraete}, which is related to the imaginary-time 
density-matrix renormalization group method~\cite{White}. The TEBD method enables us to 
access up to $N = 300$. For the detection of the phase boundary, the
Pfaffian technique is efficient, with which square-shaped systems up to 512 by 512 have been 
treated by Thomas and Katzgraber~\cite{Thomas}. 

In this article, we focus on the spin distribution function on the system boundary. If one 
regards the function as a quantum wave function, the concept of entanglement can be 
introduced~\cite{Eisert_}, in which the entanglement entropy $S$ is a typical 
measure~\cite{Srednicki}. In uniform systems, $S$ exhibits a singular behavior at 
criticality~\cite{Kitaev, Calabrese}. The presence of critical singularity in $S$ can also be 
expected in classical random systems. To confirm this conjecture, we 
numerically calculate $S$ of the square-lattice $\pm J$ Ising model.  On the Nishimori line, 
the averaged entanglement entropy $\langle S \rangle$ has a peak near the phase boundary.
By performing the finite size scaling (FSS)~\cite{Fisher, Barber}, we confirmed that the 
peak really reflects the critical singularity. Note that Ohzeki and Jacobsen 
observed a quantity that is related to the change in $S$ upon the modification 
of boundary conditions~\cite{Ohzeki_dual}. 

This article is structured as follows. In the next section, we explain the transfer matrix 
formalism in the square-lattice $\pm J$ Ising model. The entanglement entropy 
$S$ is defined through the boundary distribution function. In Sec. III, we show the numerical 
results obtained by the TEBD method. Conclusions are summarized in the last section.

\section{Model and Entanglement Entropy}

We consider the $\pm J$ Ising model on the square lattice, whose Hamiltonian is written as
\begin{equation}
H \, = \, 
\sum_{\ell, m}^{~} \, \biggl[ 
I_{\ell}^{m} \, \sigma_{\ell}^{m} \sigma_{\ell}^{m+1} +
J_{\ell}^{m} \, \sigma_{\ell}^{m} \sigma_{\ell+1}^{m} \biggr] \, ,
\end{equation}
where $\sigma_\ell^{m} = \pm 1$ denotes the Ising spin in the $\ell$-th column and 
$m$-th row. The interactions in the vertical and horizontal lattice directions are, respectively, 
denoted by $I_{\ell}^{m}$ and $J_{\ell}^{m}$. These parameters randomly take the values 
$- J < 0$ and $J > 0$, respectively, with the probability $p$ and $1 - p$. We assume that 
there is no external  field. 
Figure~1 shows the phase diagram of this 
model~\cite{Morgenstern, Hasenbusch, Honecker1, Thomas}. 
There is a ferromagnetic region when $T$ is sufficiently low and  $p$ is close to unity. 
The dashed curve denotes the Nishimori line, which is specified by~\cite{Nishimori1, Nishimori2}
\begin{equation}
\tanh \frac{J}{kT} = 2 p - 1 \, ,
\end{equation}
where $k$ denotes the Boltzmann constant. On the curve, the thermal average
of the bond energy can be exactly expressed as~\cite{Nishimori1, Nishimori2}
\begin{equation}
\langle \varepsilon \rangle = - J \tanh \frac{J}{kT} \, .
\end{equation}
The curve crosses the phase boundary at the Nishimori point $( p, T ) = 
( p_{\rm c}^{~}, T_{\rm c}^{~} )$. Below the point, the phase transition belongs to the 
percolation universality~\cite{Jacobsen}, and above the point,  it is the Ising universality. 
Note that $\langle \varepsilon \rangle$ in Eq.~(3) shows no singularity 
in any temperature. 

\begin{figure}
\begin{center}
\includegraphics[width = 4.6 cm]{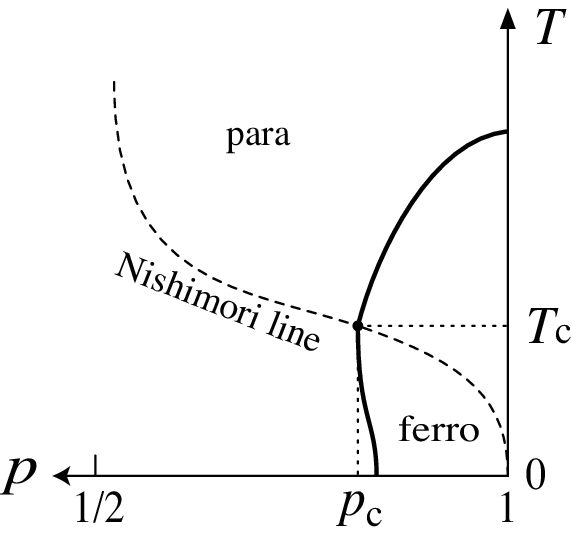}
\end{center}
\caption{Phase diagram of the $\pm J$ Ising model on the square 
lattice~\cite{Morgenstern, Hasenbusch, Honecker1, Thomas}. }
\end{figure}

We represent the system as the random interaction-round-a-face model, where
each {\it face} surrounded by $\sigma_{\ell}^{m}$, $\sigma_{\ell+1}^{m}$, $\sigma_{\ell}^{m+1}$, 
and $\sigma_{\ell+1}^{m+1}$ is considered as the unit of the system. The 
corresponding local Boltzmann weight is given by
\begin{eqnarray}
W_{\ell}^{m}  & \!\!\! = \!\!\! & \exp \biggl[ - \frac{1}{2kT} \Bigl(
I_{\ell}^{m} \sigma_{\ell}^{m} \sigma_{\ell}^{m+1} + 
I_{\ell+1}^{m} \sigma_{\ell+1}^{m} \sigma_{\ell+1}^{m+1} \biggr. \nonumber\\
&& ~~~~~~~~~~~~ \biggl. ~ + 
J_{\ell}^{m} \sigma_{\ell}^{m} \sigma_{\ell+1}^{m} + 
J_{\ell}^{m+1} \sigma_{\ell}^{m+1} \sigma_{\ell+1}^{m+1} \Bigr) \biggr] \, ,
\end{eqnarray}
where the factor $1/2$ indicates that each bond is shared by two faces. 
In the following, we consider a rectangular system of horizontal width $N$ 
and height $M$. We impose open boundary conditions for all the system boundaries. 
The partition function of the system is then represented as 
\begin{equation}
Z = \sum_{\rm conf.}^{~} \,  \prod_{m = 1}^{M-1} \, \prod_{\ell = 1}^{N-1} \, W_{\ell}^{m} \, ,
\end{equation}
where the spin configuration sum is taken over. The interaction on the system 
boundary is $\pm J/2$ by definition of $W_{\ell}^{m}$. Let us express the row 
of spins $\sigma_1^{m}, \sigma_2^{m}, \ldots$, and $\sigma_N^{m}$ by the 
notation $\{ \sigma_{~}^{m} \}$, and define the transfer matrix
\begin{equation}
U^{m}_{~}\Bigl( \, \{ \sigma_{~}^{m+1} \} \, \Big| \, \{ \sigma_{~}^{m} \} \, \Bigr) \, = \, 
\prod_{\ell = 1}^{N-1} \, W_{\ell}^{m} \, ,
\end{equation}
whose structure is shown in Fig.~2. We can then express $Z$ as a 
contraction of the product of transfer matrices
\begin{equation}
Z \, = \, \sum_{ \{ \sigma_{~}^{M} \} }^{~} \, \sum_{ \{ \sigma_{~}^{1} \} }^{~} \, 
U^{M-1}_{~} U^{M-2}_{~} \, \cdots \, U^{2}_{~} U^{1}_{~} \, .
\end{equation}
Skipping the summation for $\{ \sigma^{M}_{~} \}$, we obtain the partial sum
\begin{equation}
V\Bigl(  \, \{ \sigma^{M}_{~} \} \, \Bigr)  \, = \, \sum_{ \{ \sigma_{~}^{1} \} }^{~} \, 
U^{M-1}_{~} U^{M-2}_{~} \, \cdots \, U^{2}_{~} U^{1}_{~} \, ,
\end{equation}
which is dependent on the spin configuration $\{ \sigma^{M}_{~} \}$ at the top of the 
system. This is the (unnormalized) distribution function that we mentioned in the previous section, 
where the normalized probability of observing a particular spin configuration on this boundary 
can be written as the ratio 
\begin{equation}
P\Bigl(  \, \{ \sigma^{M}_{~} \} \, \Bigr) = \frac{1}{Z} \, 
V\Bigl(  \, \{ \sigma^{M}_{~} \} \, \Bigr) \, .
\end{equation}

\begin{figure}
\begin{center}
\includegraphics[width = 6.4 cm]{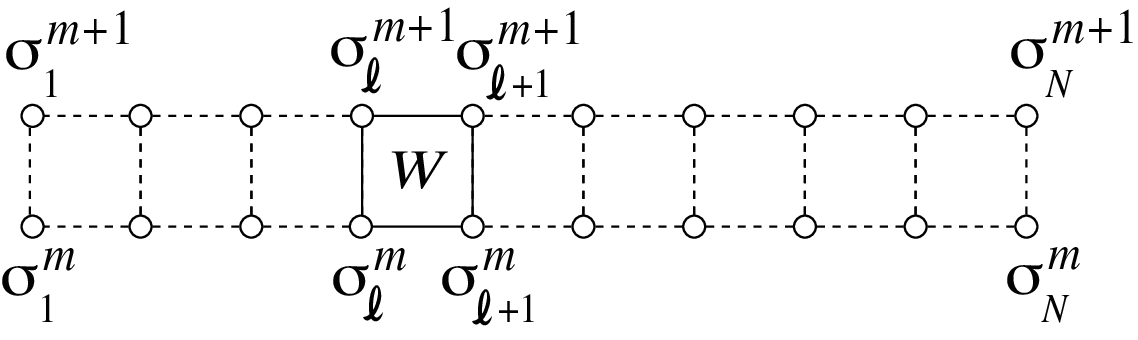}
\end{center}
\caption{Structure of the transfer matrix $U^m_{~}$.}
\end{figure}

We interpret the partial sum $V$ in Eq.~(8) as an unnormalized wave function 
of a hypothetical one-dimensional spin system. 
For later convenience, let us introduce the normalized wave function
\begin{equation}
{\Psi}\Bigl(  \, \{ \sigma^M_{~} \} \, \Bigr) = 
\frac{1}{\sqrt{ \cal N } } \, V\Bigl(  \, \{ \sigma^M_{~} \} \, \Bigr) \, ,
\end{equation}
where ${\cal N} = 
\sum_{ \{ {\sigma}^M_{~} \} }^{~} \biggl[ V\Bigl(  \, \{ {\sigma}^M_{~} \} \, \Bigr) \biggr]^2_{~}$ 
is the square of the norm. The concept of quantum entanglement can be introduced
to an arbitrary quantum state. Let us divide $\{ \sigma^M_{~} \}$ into the left half 
$\{ \sigma_{\rm L}^{~} \} \equiv \sigma_1^{M}, \ldots, \sigma_{N/2}^{M}$ and the right half 
$\{ \sigma_{\rm R}^{~} \} \equiv \sigma_{N/2+1}^{M}, \ldots, \sigma_N^{M}$. By applying the
singular value decomposition
\begin{equation}
{\Psi}\Bigl(  \, \{ \sigma_{\rm L}^{~} \}, \{ \sigma_{\rm R}^{~} \} \, \Bigr)
= \sum_{\xi}^{~} \, \lambda_{\xi}^{~} \, 
A_{\xi}^{~}\Bigl(  \, \{ \sigma_{\rm L}^{~} \}  \, \Bigr) \, 
B_{\xi}^{~}\Bigl(  \, \{ \sigma_{\rm R}^{~} \}  \, \Bigr) \, ,
\end{equation}
where $A$ and $B$ are orthogonal matrices, we obtain the singular value
$\lambda_{\xi}^{~}$, which satisfies the normalization 
$\sum_{\xi}^{~} \lambda_{\xi}^{2} = 1$. The bipartite entanglement entropy
\begin{equation}
S = - \sum_{\xi}^{~} \lambda_{\xi}^{\, 2} \, \ln \, \lambda_{\xi}^{\, 2} 
\end{equation}
is a good measure of the entanglement~\cite{Srednicki}.

In uniform systems, $S$ is asymptotically 
proportional to the logarithm of the correlation length~\cite{Kitaev, Calabrese}. 
Because of  the randomness, $S$ of the $\pm J$ Ising model is dependent on the spatial
distribution of positive and negative bonds. Instead of taking a random average directly, 
we successively obtain an ensemble of $V$ by increasing the system height $M$. Using the 
self-averaging property~\cite{Virasoro} in the $\pm J$ Ising model, 
we evaluate the average $\langle S \rangle$ numerically.

\section{Calculated Results}

The partial sum $V$ in Eq.~(8) can be obtained with high numerical precision by 
the TEBD method~\cite{Vidal, Daley, Verstraete}, where $V$ is represented in the form 
of the canonical matrix product~\cite{Ostlund, Vidal_0, Schollwoeck}. The multiplication of the 
transfer matrix $V' = U^M_{~} V$ is performed by applying each $W_{\ell}^{M}$ to 
$V$ and taking the configuration sum locally. Note that each $W_{\ell}^{M}$ does 
not represent local unitary evolution; therefore, the obtained $V'$ is not represented as 
the canonical matrix product~\cite{Schollwoeck}. We transform $V'$ 
into the canonical matrix product before we evaluate $S$. Singular values $\lambda_{\xi}^{~}$ are 
obtained naturally in the numerical calculation by the TEBD method. We treat the system 
size up to $N = 300$. The system size limitation is chiefly due to the computational time 
required for the random average, while the memory/storage requirement is not severe. 
Most of the numerical calculations are performed on the K-computer.

We choose the parameter $J$ as the unit of energy, and set $k = 1$. All the calculations 
are performed on the Nishimori line, to analyze the singularity in 
$\langle S \rangle$ at the Nishimori point. 
The necessary matrix dimension $\chi$ in the TEBD method 
is dependent on $N$. We checked the convergence in $\langle S \rangle$ with 
respect to $\chi$ for the worst case, when $N = 300$ and at the Nishimori point. 
From the trial calculations up to $\chi = 28$, it is confirmed that the $\chi$-dependence is 
negligible when $\chi \ge 22$. Thus, we choose $\chi = 24$ in the following calculations. 
The number of samples of the partial sum $V$ is chosen from $D = 5 \times 10^4_{~}$ to 
$D = 2.5 \times 10^6_{~}$, depending on the system size $N$. In the critical region, 
the effective number of independent samples is estimated as $D / N$. Thus, we divide 
the $D$ numbers of calculated $S$ into bins, each containing 1000 steps, and calculate 
the subaverages in each bin. Assuming the Gaussian distribution, we estimate the standard 
deviation $\sigma$ in $\langle S \rangle$~\cite{Sandvik}, and $2\sigma$ is 
considered as the error. 

\begin{figure}
\begin{center}
\includegraphics[width = 7.7 cm]{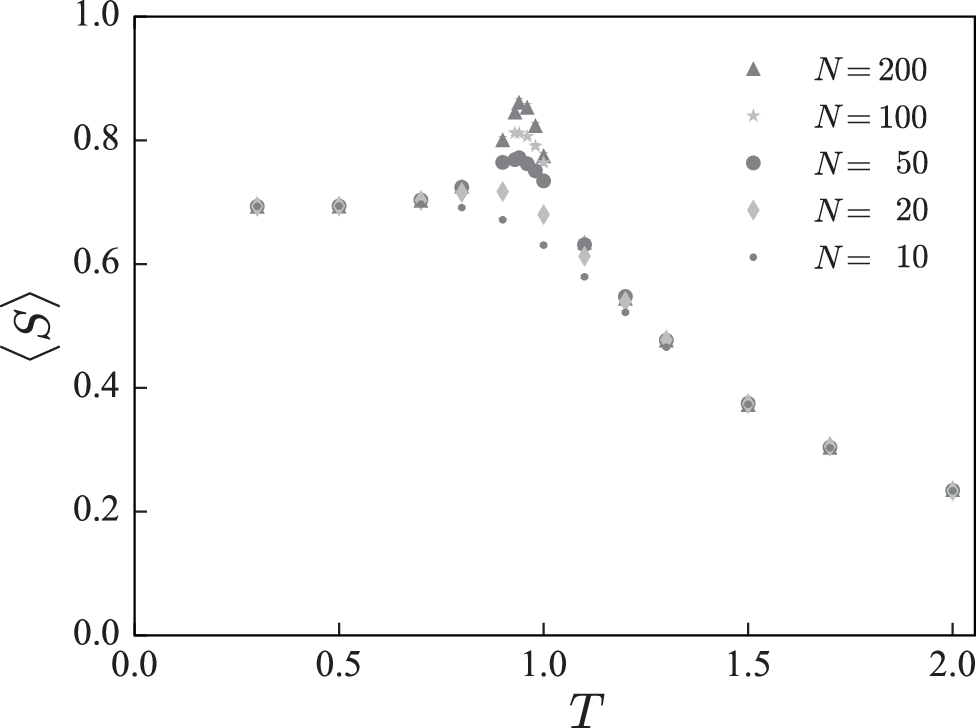}
\end{center}
\caption{Entanglement entropy $\langle S \rangle$ on the Nishimori line.}
\end{figure}

\begin{figure}
\begin{center}
\includegraphics[width = 7.9 cm]{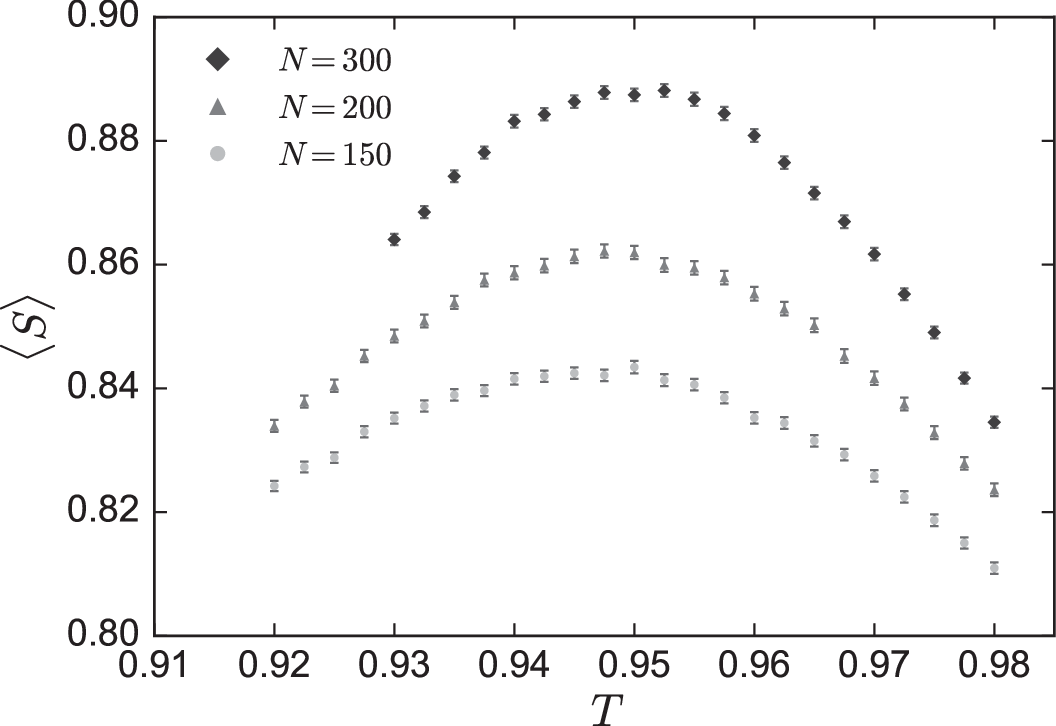}
\end{center}
\caption{Entanglement entropy $\langle S \rangle$ around $T \sim 0.95$. }
\end{figure}

Figure~3 shows the calculated $\langle S \rangle$ on the Nishimori line. Each plot is 
obtained from $D = 5 \times 10^4_{~}$ samples. When the temperature $T$ is sufficiently 
large and $p$ is close to $1/2$, $\langle S \rangle$ is a decreasing function of $T$. In the 
low-temperature limit $( p, T ) = ( 1, 0 )$, the corresponding state is the superposition 
of all-up and all-down states, which is the Greenberger--Horne--Zeilinger (GHZ) 
state~\cite{GHZ}, and $\langle S \rangle$ is equal to $\ln 2$. When the 
system size is relatively large, a peak appears in the neighborhood of $T \sim 1.0$, 
where the peak height increases with $N$. To observe the peak structure of 
$\langle S \rangle$ in detail, we calculate it within the temperature window 
$0.92 \le T \le 0.98$ as shown in Fig.~4, where $D = 1.5 \times 10^6_{~}$ samples are 
taken for the cases $N = 150$ and $200$, and $D = 2.5 \times 10^6_{~}$ for $N = 300$. 
The error bars become visible at this magnification. 

\begin{figure}
\begin{center}
\includegraphics[width = 7.9 cm]{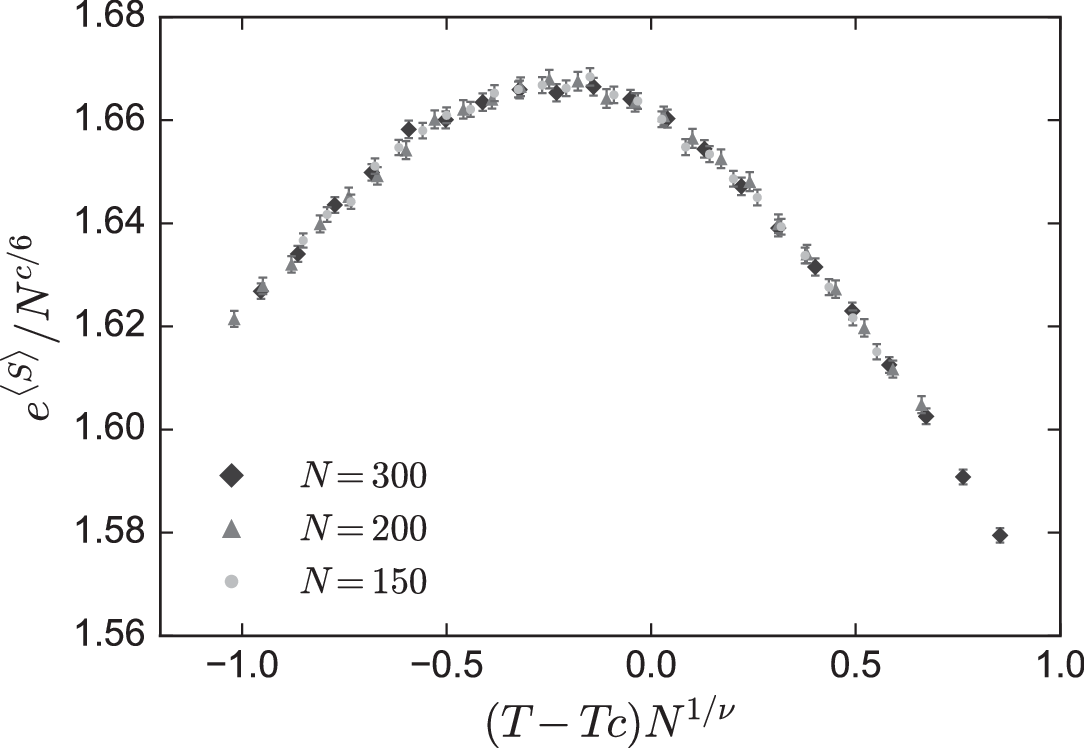}
\end{center}
\caption{Scaling plot of entanglement entropy.}
\end{figure}

We perform FSS for the plots shown in Fig.~4 for the cases $N = 150, 200$, and 
$300$, assuming the scaling form
\begin{equation}
e^{\langle S \rangle}_{~} = 
N^{c / 6}_{~} \, f\left[ \left( T - T_{\rm c}^{~} \right) N^{1 / \nu}_{~} \right] \, ,
\end{equation}
and determine the critical temperature $T_{\rm c}^{~}$ in the 
thermodynamic limit $N \to \infty$. We employ the Bayesian inference method by 
Harada~\cite{Harada}, which has also been applied to random 
systems~\cite{Nakamura,Dupont}. The temperature region $0.92 \le T \le 0.98$ 
is considered for the cases $N = 150$ and $200$, and $0.93 \le T \le 0.98$ is 
considered for $N = 300$. Figure 5 shows the obtained scaling plot. From this best fit, 
the critical temperature is estimated as $T_{\rm c}^{~} = 0.9564(3)$, where the 
corresponding probability is $p_{\rm c}^{~} = 0.89004(6)$, which is slightly smaller 
than the previously reported ones, $p_{\rm c}^{~} = 0.8906$ -- $0.8908$~\cite{Honecker2, 
Hasenbusch, Queiroz, Toldin}. As the estimation of the critical exponent, 
$\nu = 1.59(4)$ is obtained, which is larger than $\nu = 1.33$ reported by 
Picco {\it et al.}~\cite{Honecker2} for the bulk part. The central charge is 
estimated as $c = 0.397(2)$. The estimated $\nu$ contains a relatively large error, 
since it is affected by the slight change in the temperature window 
$0.92 \le T \le 0.98$, whereas the estimations for $T_{\rm c}^{~}$ and $c$ are insensitive. 

\begin{figure}
\begin{center}
\includegraphics[width = 7.9 cm]{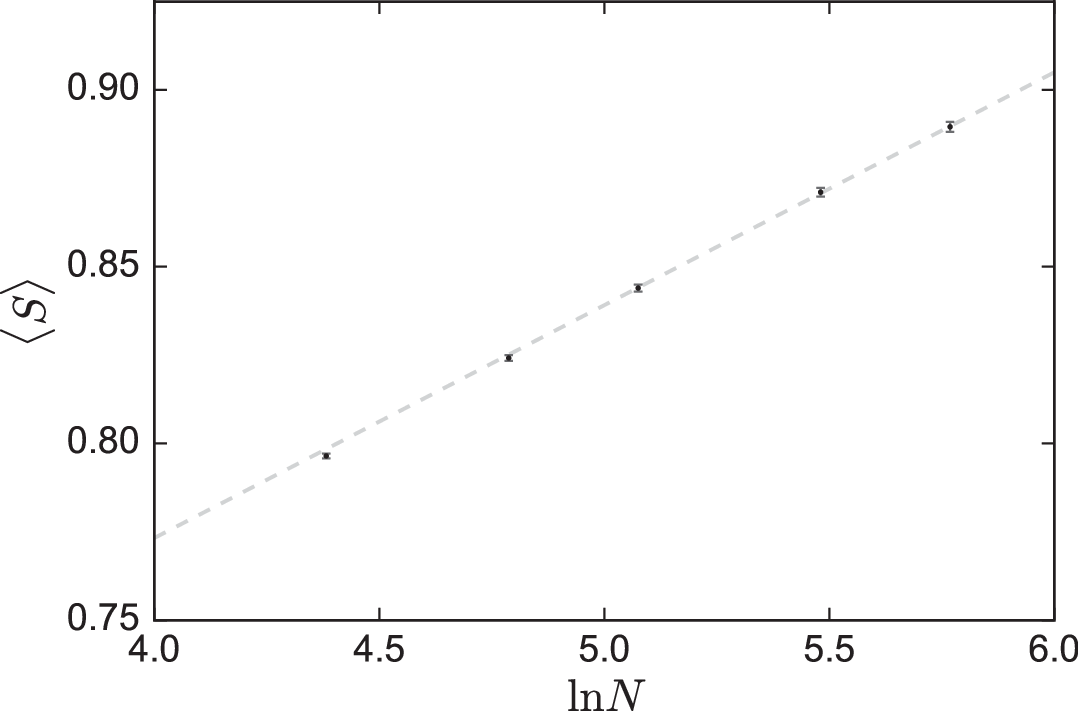}
\end{center}
\caption{Value of $\langle S \rangle$ at $T = 0.9564$ on the Nishimori line. }
\end{figure}

We perform an additional calculation at the estimated $T_{\rm c}^{~} = 0.9564(3)$, collecting 
$D = 1.5 \times 10^6_{~}$ samples for $N = 80, 120, 160, 240$, and $320$. Figure~6 
shows the obtained $\langle S \rangle$ with respect to $\ln N$. The linear dependence 
$\langle S \rangle \propto \ln N$ is clearly observed. This is in accordance with the conformal 
invariance at criticality, where the leading term of the entropy is given by 
${\displaystyle \frac{c}{6} \ln N}$. 
We obtain $c = 0.404(3)$ if we use all the plotted data in Fig.~6, 
and $c = 0.395(4)$ if we consider the cases $N = 160$, $240$, and $320$ only. The 
latter estimate is consistent with $c = 0.397(2)$ obtained from the FSS in Fig.~5. 
Our estimated $c$, which is obtained from the observation on the system boundary, 
is smaller than $c = 0.464$ estimated by Picco {\it et al.}~\cite{Honecker2} and $c = 0.463$ estimated by 
de Queiroz {\it et al.}~\cite{Queiroz} that are observed in the bulk. The discrepancy might arise from 
the difference between boundary and bulk critical phenomena.

\section{Discussion and Conclusions}

The averaged entanglement entropy $\langle S \rangle$ of the square-lattice $\pm J$ Ising model 
is analyzed through the observation of the distribution function on the system 
boundary by the transfer matrix formalism combined with the TEBD method. 
From the FSS analysis applied to the calculated $\langle S \rangle$ on the Nishimori line, the
presence of a singular behavior in $\langle S \rangle$ is confirmed around the Nishimori 
point. 

The behavior of $\langle S \rangle$ on the phase boundary apart from the Nishimori
point is a remaining point of interest. We performed small-scale trial calculations. On the phase 
boundary between the Nishimori point and the transition point of the pure Ising model with 
$p = 1$, the calculated result agrees with the Ising universality. The numerical analysis 
below the Nishimori point is not straightforward, since the spontaneous symmetry breaking 
easily occurs when the ferromagnetic bonds are accidentally concentrated. Stabilization 
should be introduced to the TEBD calculation in this case. 

Another point of interest is in the spatial structure of the entanglement on the system boundary. 
Analysis of such a structure has been carried out for one-dimensional random-bond quantum spin 
chains, which have a layered structure in entangled pairs~\cite{Ruggiero, Alba}. In the case 
of the square-lattice $\pm J$ Ising model, the randomness is present in  both horizontal 
and vertical directions of the lattice. A method of treating such disorder is the use of the tensor 
renormalization group (TRG)~\cite{Levin, Xiang}, which was once applied to the $\pm J$ 
Ising model~\cite{Guven, Wang}. From the viewpoint of the modern tensor network 
renormalization (TNR) formalisms~\cite{Evenbly, Loop, Frank}, capturing the entanglement 
structure contained in the system is essential for numerical renormalization-group 
transformations. What would be the appropriate, or adaptive, tensor-network structure 
under such randomness?

The $\pm J$ Ising model on the square lattice does not possess the spin glass phase; 
therefore, it is not possible to observe singular behaviors of the entanglement entropy around 
the spin-glass transition. Such a study can be performed on the cubic lattice, whereas  
the application of the TEBD method would require extensive computation, 
which could be undertaken by the next generation of high-performance computers.

\begin{acknowledgment}

Y.S. is grateful to Sysmex Corporation for financial support and continuous encouragement. 
This research was supported by MEXT as ``Exploratory Challenge on Post-K computer''
(Frontiers of Basic Science: Challenging the Limits).
J.G. and T.N. were supported by JSPS KAKENHI Grant Numbers 17K05578 and P17750. 
H.U. was supported by JSPS KAKENHI Grant Numbers 25800221 and 17K14359, 
and by JST PRESTO No. JPMJPR1911.
A.G. acknowledges the support by project EXSES APVV-16-0186.

\end{acknowledgment}


\begin{thebibliography}{99}
\bibitem{Edwards} S.~F.~Edwards and P.~W.~Anderson, J. Phys. F {\bf 5}, 965 (1975).
\bibitem{Cannela} V.~Cannella and J.~A.~Mydosh, Phys. Rev. B {\bf 6}, 4220 (1972).
\bibitem{Morgenstern} I.~Morgenstern and K.~Binder, Phys. Rev. B {\bf 22}, 288 (1980).
\bibitem{Hasenbusch} M.~Hasenbusch, F.~P.~Toldin, A.~Pelissetto, and E.~Vicari, 
Phys. Rev. E {\bf 77}, 051115 (2008).
\bibitem{Bhatt} R.~N.~Bhatt and A.~P.~Young, Phys. Rev. B {\bf 37}, 5606 (1988).
\bibitem{Nishimori1} H.~Nishimori, J. Phys. C {\bf 13}, 4071 (1980).
\bibitem{Nishimori2} H.~Nishimori, Prog. Theor. Phys. {\bf 66}, 1169 (1981).
\bibitem{Ozeki} Y.~Ozeki and H.~Nishimori, J. Phys. Soc. Jpn. {\bf 56}, 3265 (1987).
\bibitem{Aarao} F.~D.~A.~Aar\~ao Reis, S.~L.~A.~de Queiroz, and R.~R.~dos Santos, Phys. Rev. B {\bf 60}, 6740 (1999).
\bibitem{Honecker1} A.~Honecker, M.~Picco, and P.~Pujol, Phys. Rev. Lett. {\bf 87}, 047201 (2001).
\bibitem{Merz} F.~Merz and J.~T.~Chalker, Phys. Rev. B {\bf 65}, 054425 (2002).
\bibitem{Honecker2} M.~Picco, A.~Honecker, and P.~Pujol, J. Stat. Mech.: Theory and Exp. {\bf 2004}, P09006 (2006).
\bibitem{Vidal} G.~Vidal, Phys. Rev. Lett. {\bf 93}, 040502 (2004).
\bibitem{Daley} A.~J.~Daley, C.~Kollath, U.~Schollw\"ock, and G.~Vidal, 
J. Stat. Mech.: Theory and Exp. P04005 (2004).
\bibitem{Verstraete} F.~Verstraete, J.~J.~Garcia-Ripoll, and J.~I.~Cirac, 
Phys. Rev. Lett. {\bf 93}, 207204 (2004). 
\bibitem{White} S.~R.~White and A.~E.~Feiguin, Phys. Rev. Lett. {\bf 93}, 076401 (2004).
\bibitem{Thomas} C.~K.~Thomas and H.~G. Katzgraber, Phys. Rev. E {\bf 84}, 040101(R) (2011).
\bibitem{Eisert_} J.~Eisert, M.~Cramer, and M.~B.~Plenio, Rev. Mod. Phys. {\bf 82}, 277 (2010).
\bibitem{Srednicki} M.~Srednicki, Phys. Rev. Lett. {\bf 71}, 666 (1993).
\bibitem{Kitaev} G.~Vidal, J.~I.~Latorre, E.~Rico, and A.~Kitaev, Phys. Rev. Lett. {\bf 90}, 227902 (2003).
\bibitem{Calabrese} P.~Calabrese and J.~Cardy, J. Phys. A {\bf 42}, 504005 (2009).
\bibitem{Fisher} M.~E.~Fisher, in {\it Proc. Int. School of Physics ``Enrico Fermi''}, 
ed. M.~S. Green (Academic Press, New York, 1971) Vol. {\bf 51}, p. 1.
\bibitem{Barber} M.~N. Barber, in {\it Phase Ttransitions and Critical Phenomena}, 
eds. C.~Domb and J.~L.~Lebowitz (Academic Press, New York, 1983) Vol. {\bf 8}, p. 146.
\bibitem{Ohzeki_dual} M.~Ohzeki and J.~K.~Jacobsen, J. Phys. A: Math. Theor. {\bf 48}, 095001 (2015).
\bibitem{Jacobsen} J.~L.~Jacobsen and J.~Cardy, Nucl. Phys. B {\bf 515}, 701 (1998).
\bibitem{Virasoro} M.~M\'ezard, G.~Parisi, and M.~Virasoro, {\it Spin Glass Theory and Beyond}, 
(World Scientific, Singapore, 1987).
\bibitem{Ostlund} S.~\"Ostlund and S.~Rommer, Phys. Rev. Lett. {\bf 75}, 3537 (1995).
\bibitem{Vidal_0} G.~Vidal, Phys. Rev. Lett. {\bf 91}, 147902 (2003).
\bibitem{Schollwoeck} U.~Schollw\"ock, Ann. Phys. {\bf 326}, 96 (2011).
\bibitem{Sandvik} The numerical error is estimated using Eq.~(48) of the following article. 
A.~W.~Sandvik, AIP Conf. Proc. {\bf 1297}:135,2010; arXiv:1101.3281.
\bibitem{GHZ} D.~M.~Greenberger, M.~A.~Horne, and A.~Zeilinger, in {\it Bell's Theorem, Quantum Theory, 
and Conceptions of the Universe}, ed. M.~Kafatos (Kluwer, Dordrecht, 1989) p. 69.
\bibitem{Harada} K.~Harada, Phys. Rev. E {\bf 84}, 056704 (2011).
\bibitem{Nakamura} T.~Nakamura and T.~Shirakura, J. Phys. Soc. Jpn. {\bf 84}, 013701 (2015).
\bibitem{Dupont} M.~Dupont, S.~I.~Capponi, and N.~Laflorencie, Phys. Rev. Lett. {\bf 118}, 067204 (2017).
\bibitem{Queiroz} S.~L.~A. de Queiroz, Phys. Rev. B {\bf 79}, 174408 (2009).
\bibitem{Toldin} F.~P.~Toldin, A.~Pelissetto, and E.~Vicari, J. Stat. Phys. {\bf 135}, 1039 (2009).
\bibitem{Ruggiero} P.~Ruggiero, V.~Alba, and P.~Calabrese, Phys. Rev. B {\bf 94}, 035152 (2016).
\bibitem{Alba} V.~Alba, S.~N.~Santalla, P.~Ruggiero, J.~Rodriguez-Laguna, P.~Calabrese, and G.~Sierra, 
J. Stat. Mech.: Theory and Exp. {\bf 2019}, 023105 (2019).
\bibitem{Levin} M.~Levin and C.~P.~Nave, Phys. Rev. Lett. {\bf 99}, 120601 (2007).
\bibitem{Xiang} Z.~Y.~Xie, J.~Chen, M.~P.~Qin, J.~W.~Zhu, L.~P.~Yang, and T.~Xiang, 
Phys. Rev. B {\bf 86}, 045139 (2012).
\bibitem{Guven} C.~G\"uven, M.~Hinczewski, and A.~N.~Berker, Phys. Rev. E {\bf 82}, 051110 (2010).
\bibitem{Wang} C.~Wang, S.~M.~Qin, and H.~J.~Zhou, Phys. Rev. B {\bf 90}, 174201 (2014).
\bibitem{Evenbly} G.~Evenbly and G.~Vidal, Phys. Rev. Lett. {\bf 115}, 180405 (2015).
\bibitem{Loop} S.~Yang, Z.~C.~Gu, and X.~G.~Wen, Phys. Rev. Lett. {\bf 118}, 110504 (2017).
\bibitem{Frank} M.~Bal, M.~Mari\"en, J.~Haegeman, and F.~Verstraete, Phys. Rev. Lett. {\bf 118}, 
250602 (2017).
\end{thebibliography}
\end{document}